\documentclass[aps,prl,superscriptaddress,reprint]{revtex4-1}

\usepackage{graphicx,multirow}
\usepackage{bm}
\usepackage{amsmath}
\usepackage{amssymb}
\usepackage{amscd}
\usepackage{latexsym}
\usepackage{slashed}
\usepackage{color}
\usepackage{graphicx}
\usepackage{hyperref}
\usepackage{ulem}
\usepackage{color}
\usepackage[caption=false]{subfig}

\def\ltap{\raisebox{-.6ex}{\rlap{$\,\sim\,$}} \raisebox{.4ex}{$\,<\,$}} 
\def\gtap{\raisebox{-.6ex}{\rlap{$\,\sim\,$}} \raisebox{.4ex}{$\,>\,$}}

\begin{document}

\title{
{TMD Evolution and Multi-Jet Merging}
}

\author{A. Bermudez Martinez}%
 \email{armando.bermudez.martinez@desy.de}
\affiliation{DESY, D-22607 Hamburg}%
\author{F.~Hautmann}%
 \email{hautmann@thphys.ox.ac.uk}
\affiliation{Universiteit Antwerpen, Elementaire Deeltjes Fysica, B 2020 Antwerpen}
\affiliation{University of Oxford, Theoretical Physics Department, Oxford OX1 3PU}%
\author{M.L.~Mangano}%
\email{michelangelo.mangano@cern.ch}
\affiliation{CERN, Theoretical Physics Department, CH 1211 Geneva}

\begin{abstract}
\noindent
The theoretical description of the physics of multi-jets in hadronic collisions at high energies is based on ``merging" 
methods, which 
combine short-timescale production of jets with long-timescale evolution of partonic showers. 
We point out potential implications of  the evolution of   
transverse momentum dependent (TMD) distributions on the structure of multi-jet states at high energies, and in particular 
on the theoretical systematics associated with multi-jet merging. 
To analyze this, we propose a new merging methodology,  and illustrate its impact by comparing  
our theoretical results with 
experimental measurements for $Z$-boson + jets production at the Large Hadron Collider (LHC). 
\end{abstract}

\maketitle

Some of the most important production channels  in high-energy collisions at the LHC and at future colliders involve final states with large multiplicities of jets.  
Such final states are used both to perform precision measurements within the Standard Model (SM) and to search for possible signatures of physics  beyond the SM (BSM).  

Theoretical predictions for multi-jet observables have relied for the last twenty years on 
merging methods to combine ``matrix-element" and ``parton-shower" contributions. The former describe the underlying hard process, at momentum transfers large compared to the Quantum Chromodynamics (QCD) mass scale $\Lambda_{\rm{QCD}}$, with bare partons (i.e., quarks and gluons) providing the primary sources for widely separated jets;  the latter 
describe the evolution of partons by radiative processes, predominantly in the collinear and soft regions. The two contributions are then sewn together via a ``merging" scheme, to prevent the double counting, or exclusion, of multi-jet phase space volumes. Merging algorithms have been developed and extensively used initially at  
leading order 
(LO)~\cite{Catani:2001cc,Lonnblad:2001iq,Mangano:2002,Mrenna:2003if,Alwall:2007fs,Mangano:2006rw,Lavesson:2008ah,Hoeche:2009rj,Hamilton:2009ne,Lonnblad:2011xx}
and later at next-to-leading order (NLO)~\cite{Frederix:2012ps,Hoeche:2012yf,Lonnblad:2012ix,Bellm:2017ktr}. 

Collinear  distributions of initial-state partons lead to a strong reduction of information and tell us only about the longitudinal momentum of partons in a fast moving hadron. 
This restriction is lifted in  the transverse momentum dependent (TMD) parton distribution functions (PDFs) ---  more general distributions which provide 
``3-dimensional (3D) imaging" of the hadron structure~\cite{Angeles-Martinez:2015sea} --- and corresponding initial-sate parton showers.  TMD 
distributions are used  to obtain QCD factorization and resummation formulas for collider  observables in kinematic regions characterized by multiple momentum scales, e.g., in the Sudakov region~\cite{Collins:1984kg} and high-energy region~\cite{Catani:1990eg}.   In the last few years their phenomenological importance has been emphasized in particular for the low transverse momentum region in Drell-Yan (DY) lepton-pair production --- see e.g.~the recent CMS experimental study~\cite{cmspas:003-2020} --- and  in
deep inelastic scattering (DIS)~\cite{Agostini:2020fmq,Aidala:2020eah}. 
For a recent study of the interplay between perturbative and non-perturbative effects induced by TMD evolution in the transverse momentum Drell-Yan spectrum at small $k_T$, see
e.g.~\cite{Hautmann:2020cyp}.

The impact of TMD distributions on the high transverse momentum region and on multi-jet production, on the other hand, is as yet unexplored. This is likely due to the prejudice that the 3D picture of the proton provided by TMD distributions plays a role only at scales of few-GeV. On the other hand, as we shall recall below, the renormalization-group evolution of TMDs to large scales induces in a natural way large, perturbative, transverse momentum tails, which impact the description of multi-jet final states. 

For a multi-jet final state characterized by the hard momentum-transfer scale $\mu$, we 
analyze the contribution to the production of an extra jet with transverse momentum 
$p_T <  \mu$ from  the high-$k_T$ tail of the initial state parton distribution, $k_T \gtap p_T$. While the distribution 
is falling off at large $k_T$, we find that for the jet transverse scales observed at the LHC  the contribution from the 
region $p_T \ltap k_T \ltap \mu$ is non-negligible when compared to that of an extra parton perturbatively emitted through hard-scattering matrix elements. 
As a result, a merging procedure is needed, to avoid the double counting between the extra jet emission induced by the TMD initial-state evolution, and that arising from the inclusion of a higher-order matrix element. 

In this letter we present such a procedure, dubbed ``TMD merging", which extends to the case of TMD initial-state evolution the familiar MLM merging procedure~\cite{Mangano:2002,Alwall:2007fs,Mangano:2006rw}. We show a few first applications, and comparisons with experimental data~\cite{Aad:2015auj,Aaboud:2017hbk}, for $Z$-boson + jets production at the LHC. 
As a main result, we find that, with respect to the standard MLM procedure, TMD merging (i) has reduced systematic uncertainties, and (ii) improves the description of higher-order emissions beyond the maximum parton multiplicity considered in the matrix element calculations. 
A more detailed account will appear in a separate 
publication~\cite{prepa21}. 

We start by recalling the parton branching formulation~\cite{Hautmann:2017xtx,Hautmann:2017fcj} of TMD evolution. 
This formulation is well-suited for investigating issues of merging in multi-jets, as it uses  at TMD level 
 the unitarity picture of parton 
evolution~\cite{Webber:1986mc,eswbook} commonly employed at the collinear level in 
showering algorithms~\cite{Bellm:2015jjp,Sjostrand:2014zea}. 
Soft gluon emission and transverse momentum recoils are treated 
by introducing the soft-gluon resolution scale    
$z_M$~\cite{Hautmann:2017xtx}  
to separate 
resolvable branchings and non-resolvable branchings, with the  former being described 
through  real-emission splitting functions and the latter through Sudakov form 
factors~\cite{Hautmann:2017fcj}. In this approach the  
TMD evolution equations are written in the schematic form 
\cite{Hautmann:2017fcj}   
\begin{eqnarray}
\label{eq1pb-tmdevol}
&& {A}_j\left( x, {\bm k}^2, \mu^2\right) = 
 \Delta_j\left(\mu^2, \mu_0^{2}\right)
 {A}_j\left( x, {\bm k}, \mu_0^2\right)
  \\ 
 &+& 
 \sum_\ell\int \frac{\textrm{d}^2{\boldsymbol \mu}^{\prime}}{\pi {\mu}^{\prime 2}} 
   \int  \textrm{d}z 
\   {\cal K}_{j \ell}    \left(  x, {\bm k}, \mu^2 ;   z , z_M ,  {\mu}^{\prime 2}     \right) 
\nonumber \\
 &\times&  {A}_\ell\left( x / z , | {\bm k} + (1-z){\boldsymbol \mu}^\prime |^2, \mu^{\prime 2}\right) \; , 
\nonumber 
\end{eqnarray}  
where: ${A}_j( x, {\bm k}^2, \mu^2)$ is the 
TMD distribution of flavor $j$ carrying the longitudinal momentum  fraction $x$ of the hadron's momentum and  transverse momentum ${\bm k}$
at the evolution scale $\mu$;  $ \Delta_j $ is the Sudakov form factor,  and $ {\cal K}_{j \ell } $    are evolution kernels, 
computable in terms of Sudakov form factors,   real-emission splitting functions and phase-space constraints taking into account soft-gluon angular  ordering~\cite{Hautmann:2019biw,Catani:1990rr,Marchesini:1987cf}. 

Eq.~(\ref{eq1pb-tmdevol}) can be viewed as a ``forward evolution" equation, in which  
  $\mu_0$ is the initial evolution scale, and 
 $z$ and ${\boldsymbol \mu}^\prime$ are the branching variables, with $z$ being the longitudinal momentum transfer  
at the  branching, and  $ \mu^\prime = \sqrt{ {\boldsymbol \mu}^{\prime 2}}$ the momentum scale at which the branching occurs. 
Once the TMD distribution $ {A}_j( x, {\bm k}^2, \mu^2)$ evaluated at the scale 
$\mu^{2}$ is known,  the corresponding TMD parton shower can be 
generated by ``backward evolution"~\cite{Baranov:2021uol}. 

The above formalism has been used, in the spirit of the renormalization group evolution for PDFs,  
to  extract TMD distributions~\cite{Martinez:2018jxt} from fits to precision DIS data, using the QCD fit platform   \verb+xFitter+~\cite{Alekhin:2014irh} (for  
 other available TMD fits, see  the library~\cite{Abdulov:2021ivr,Hautmann:2014kza}). Furthermore, the formalism has been 
 used to make predictions for the DY $p_T$ spectrum~\cite{Martinez:2019mwt},   
 including Sudakov resummation through next-to-leading-logarithmic accuracy and  
 matching with 
 next-to-leading-order (NLO) matrix elements~\cite{Alwall:2014hca}.  
A good description of 
 DY measurements is achieved across a wide range of energies and masses,  from the LHC down to 
 fixed-target experiments~\cite{Martinez:2019mwt,Martinez:2020fzs}. 
The physical picture emerging from the above studies is that TMD distributions are 
characterized by  transverse momentum ($k_T = | {\bm k} |$) widths ($\sigma$) of the order of 
$\Lambda_{\rm{QCD}} \ltap \sigma \ltap 1 \ {\rm GeV}$ at $\mu_0 \sim {\cal O} ( 1 \  {\rm GeV})$, and undergo 
$k_T$ broadening as the evolution scale $\mu$ increases due to the interplay of 
 resolvable and non-resolvable branchings with the initial-scale distribution.

\begin{figure}[hbtp]
  \begin{center}
	\includegraphics[width=.40\textwidth]{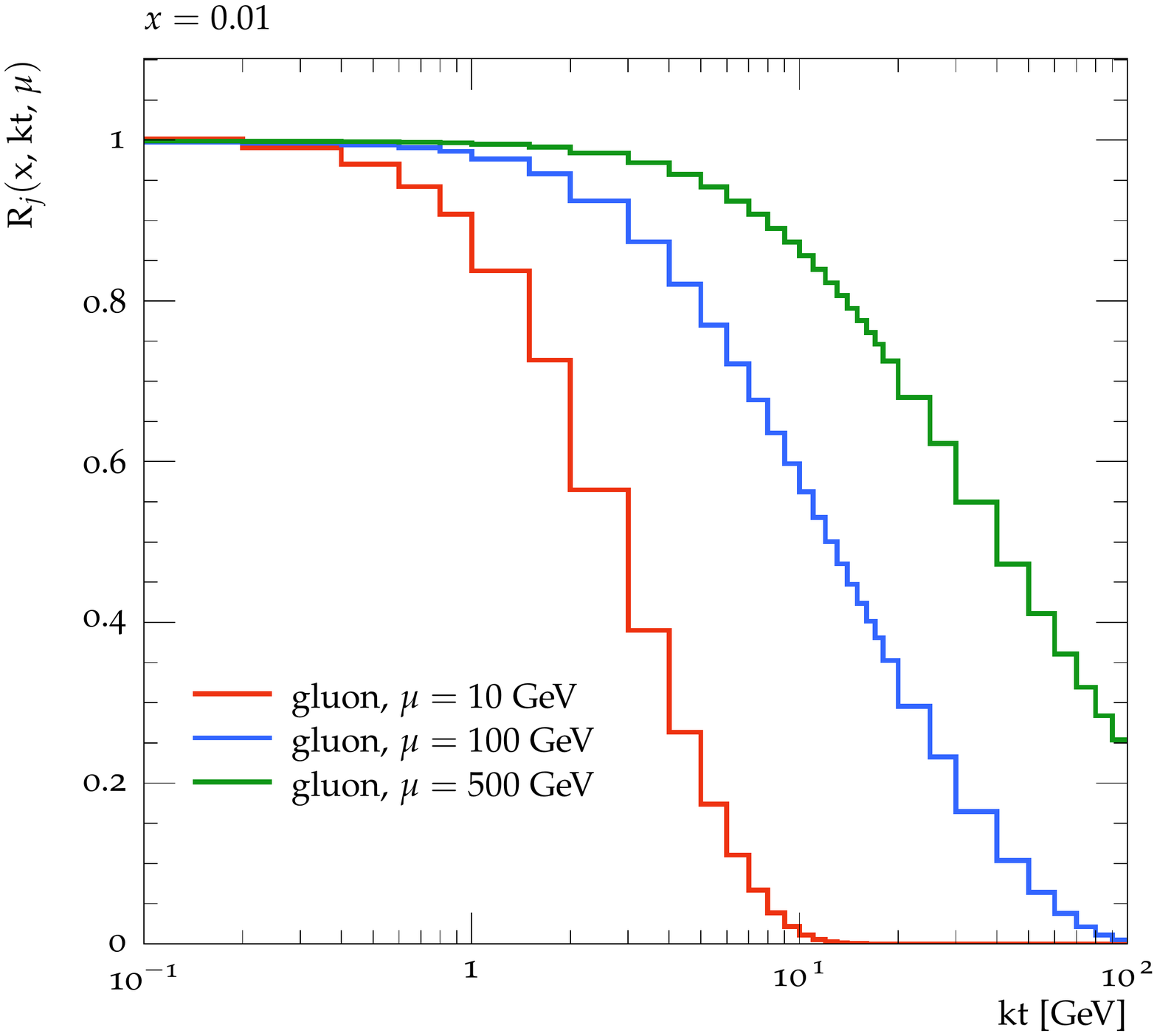}
  \caption{The $k_T$ spectrum of the integral  
  TMD  gluon distribution, normalized to $k_T=0$  as   in Eq.~(\ref{normalized-itmd}),  
  for longitudinal momentum fraction $x= 10^{-2}$ and different values of the evolution scale $\mu$. 
  The PB TMD Set~2~\cite{Martinez:2018jxt} is used.}
\label{fig1:iTMD-vs-kTmin}
\end{center}
\end{figure}

Let us then consider a final state in high-energy hadronic collisions characterized 
by a hard scale $\mu$, e.g. the transverse momentum of the hardest jet in the event. 
What is the contribution to the emission of an extra jet of transverse momentum $p_T$, with $p_T$ lower than $\mu$, from the  $k_T$ broadening of the TMD distribution evolved to scale $\mu$? To estimate this,  we introduce integral TMD distributions $a_j$, obtained from  
${A}_j$ in Eq.~(\ref{eq1pb-tmdevol}) by $k_T$-integration as follows 
 \begin{equation} 
\label{eq2integraltmd}
a_j ( x, {\bm k}^2, \mu^2) = 
\int    { {d^2 {\bm k }^\prime} \over \pi} \  {\cal A}_j ( x , {\bm k }^{\prime 2} , \mu^2) \  \Theta ( {\bm k }^{\prime 2}-{\bm k }^{ 2} ) 
\; . 
\end{equation} 
The distribution $a_j$ evaluated at $k_T = 0$ gives the fully integrated initial-state distribution, namely a standard collinear PDF. 
We are interested in the fractional contribution to $a_j$ from the tail above transverse momentum $k_T$, with $k_T$ of the order of 
the jet $p_T$. For any flavor $j$ we thus construct the ratio 
\begin{equation}
\label{normalized-itmd} 
R_j ( x, {\bm k}^2, \mu^2) = a_j ( x, {\bm k}^2, \mu^2) / a_j ( x, { 0}, \mu^2)  \; .  
\end{equation} 
In Fig.~\ref{fig1:iTMD-vs-kTmin} we show the $k_T$ dependence of Eq.~(\ref{normalized-itmd})   
by an example showing the integral TMD gluon distribution $a_g ( x, {\bm k}^2, \mu^2) $, normalized to    $k_T=0$,  
 obtained from the TMD fitted to precision DIS data in~\cite{Martinez:2018jxt} (PB TMD Set~2), for $x=10^{-2}$ and various values of $\mu$. We observe, for 
 instance, that for $\mu=100$ (500) GeV, there is a 30\% probability that the gluon has developed a transverse momentum larger than 20 (80) GeV. 
 
The results in Fig.~\ref{fig1:iTMD-vs-kTmin} confirm that the $k_T$ tail contribution to jet emission is comparable, in the LHC kinematics, to perturbative emissions via hard matrix elements. 

\begin{figure}[hbtp]
  \begin{center}
	\includegraphics[width=.40\textwidth]{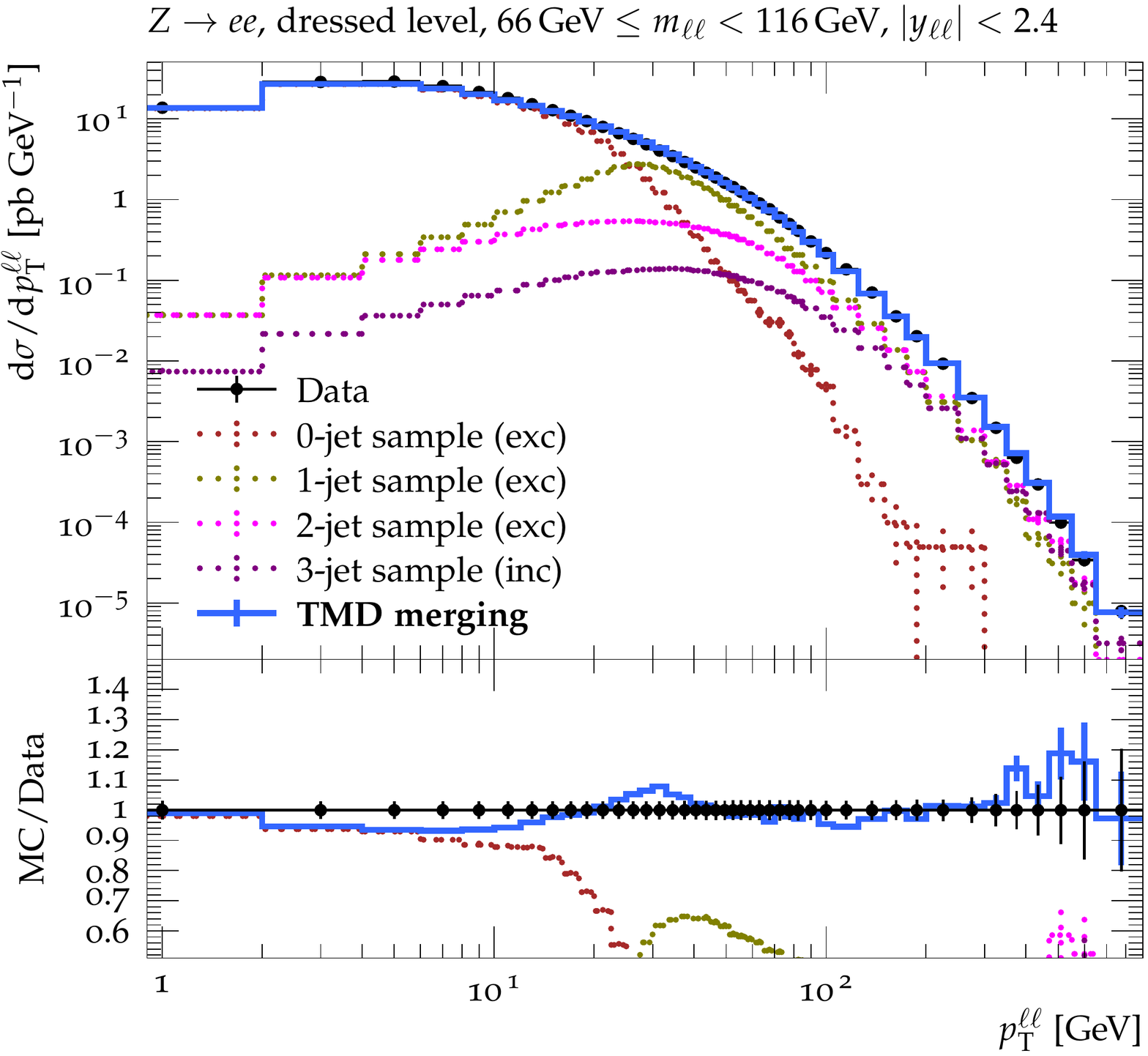}
  \caption{Transverse momentum $p_T$ spectrum of  DY lepton pairs from $Z$-boson decays. Experimental 
  measurements by ATLAS~\protect\cite{Aad:2015auj} at 
  $\sqrt{s} = 8$ TeV are shown. The result of the 
  fully TMD-merged calculation, as well as separate contributions from 
  the different jet samples, are shown. All jet multiplicities are obtained in exclusive (exc) mode except for the 
  highest multiplicity which is calculated in 
  inclusive (inc) mode.}
  \label{fig2:ZpT}
  \end{center}
\end{figure}

The TMD merging approach complements existing approaches, which  rely 
on merging samples of different parton multiplicity showered through 
emissions in the collinear approximation, with the use of the TMD parton 
branching for the initial state evolution. 
To begin our exploration, we shall work at the LO level, expanding on the  
MLM merging approach~\cite{Mangano:2002,Mrenna:2003if,Alwall:2007fs,Mangano:2006rw}. We expect a similar construction to be possible starting from other approaches, such as CKKW-L~\cite{Catani:2001cc,Lonnblad:2001iq}. 

The key features of TMD merging, with respect to MLM, are summarized here (for more details, see~\cite{prepa21}). For each n-jet parton-level event, transverse momentum vectors ${\bm k_{1,2}}$ are selected for the two initial-state partons with longitudinal momentum $x_{1,2}$. The values of ${\bm k_{1,2}}$ are extracted from the ${\bm k}^2$ distribution defined by the solution of eq.~\ref{eq1pb-tmdevol} for ${A}_j\left( x_{1,2}, {\bm k}^2, \mu^2\right)$. If ${\bm k_1^2}$ or ${\bm k_2^2}$ are larger than some cutoff $\mu^2_{min}$, the event is rejected. We define $\mu^2_{min}={\mbox{min}}\{ p^2_{ti}, p^2_{tij} \}$ where $i,j=1,..,n$,  $p_{ti}$ is the transverse momentum of  parton $i$, and $p^2_{tij}$ ($i\neq j$) measures the relative transverse momentum between partons $i$, $j$. This rejection plays the role of a Sudakov re-weighting, accounting for the suppression of hard initial-state radiation of jets harder than those already present in the matrix-element calculation. The transverse vector $\bm{k=k_1+k_2}$ is applied as a boost to the final partonic state. This event is then showered, replacing the standard initial-state backward evolution with the backward evolution driven by the TMD evolution of Eq.~(\ref{eq1pb-tmdevol}).   

The usual final part of MLM merging is eventually applied, matching the directions of the showered jets to those of the initial partons (boosted by $\bm k$), and cutting on the possible emission of additional jets. In particular, if the maximum jet multiplicity for which matrix elements are calculated is $N_{max}$, MLM merging requires that the total number of reconstructed jets equals the number of partons for all $n$-parton samples up to $n=N_{max}-1$ (``exclusive" multiplicity samples), while extra jets are allowed when $n=N_{max}$, provided they are softer than the jets matched to the original partons (``inclusive" multiplicity sample)~\cite{footnote1}. 

As an illustration, we consider the case of $Z$-boson production in association with jets at the LHC, at  center-of-mass 
energies $\sqrt{s} =8$ and 13~TeV. For this study, we use {\sc MadGraph5\_aMC@NLO}~\cite{Alwall:2014hca} to generate $Z$ $+0,1,2,3$ jet samples 
at LO with a partonic generation cut $p_T>15$ GeV. We use the event generator 
 {\sc Cascade}~\cite{Baranov:2021uol,Jung:2010si} to generate the TMD backward shower, and {\sc Pythia}6.4~\cite{Sjostrand:2006za} 
 for the final-state shower.  We apply the  parton distributions 
 obtained from DIS fits in~\cite{Martinez:2018jxt} with $\alpha_s(M_Z) = 0.118$. The nominal value for the MLM merging scale 
 is chosen to be $\mu_{m}= 23$ GeV.
Following~\cite{Mrenna:2003if,Alwall:2007fs},   we have tested the consistency and systematics of TMD merging   
by computing the differential jet rates~\cite{prepa21} in the variable $d_{n,n+1}$ (for $n = 0, 1, 2$), the square of the energy scale at which an 
$n$-jet event is resolved as an $(n+1)$-jet event, with parton-level jets reconstructed by the 
$k_\text{t}$ jet-clustering~\cite{Catani:1993hr,Ellis:1993tq}.

In Fig.~\ref{fig2:ZpT} we 
show the transverse momentum spectrum of DY lepton pairs from $Z$-boson decay in $pp$ collisions at $\sqrt{s} = 8$ TeV, 
and compare the results with the ATLAS measurements~\cite{Aad:2015auj}. 
The analysis is performed using   {\sc Rivet}~\cite{Buckley:2010ar}.  
The distribution is normalized to the next-to-next-to-leading-order (NNLO) DY cross section.   The contributions from the different jet multiplicities 
to the final prediction are shown separately.            

We observe that the $Z$+0 jet sample constitutes the main contribution at low transverse momentum $p_T$ while the impact of larger jet multiplicities gradually increases with 
increasing $p_T$. The merged prediction provides a good description of the data, with an agreement within $\pm 10\%$ throughout the whole DY $p_T$ spectrum. 

\begin{figure}[hbtp]
  \begin{center}
	\includegraphics[width=.40\textwidth]{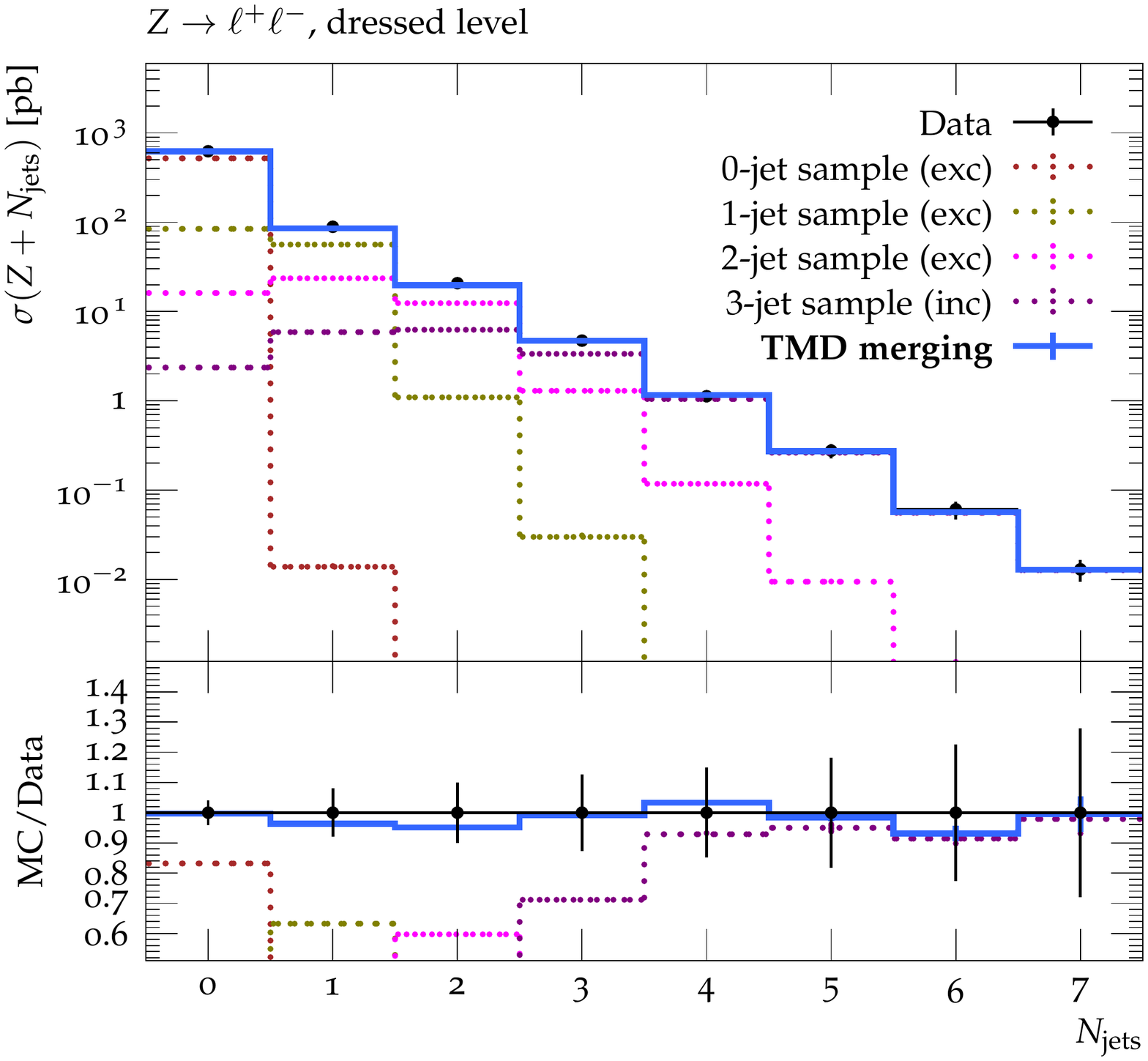}
  \caption{Jet multiplicity in the production of a $Z$-boson in association with jets. 
  Experimental measurements by ATLAS~\protect\cite{Aaboud:2017hbk} at  $\sqrt{s} = 13$ TeV are shown for comparison. We plot the results of the 
     TMD merged calculation, and the inidividual contributions from the different jet samples. In the legend, (inc) and (exc) refer 
     to the inclusive or exclusive multiplicity definition of a given $n$-jet sample, as discussed in the text.}
  \label{fig3:jetmult}
  \end{center}
\end{figure}

Figure~\ref{fig3:jetmult} shows our results for  the jet multiplicity distribution in $Z$+jets production at  $\sqrt{s} = 13$  TeV. As in the case of 
Fig.~\ref{fig2:ZpT}, the result is normalized to the NNLO DY cross section, and is compared to the ATLAS measurement~\cite{Aaboud:2017hbk} (similar results have been reported by the CMS experiment~\cite{Sirunyan:2018cpw}). 
Jets are defined by the anti-$k_t$ algorithm~\cite{Cacciari:2008gp} with radius $R=0.4$, and are required to have $p_T>30$~GeV and $\vert \eta\vert<2.5$.

The agreement of the prediction with the experimental measurements shown in Fig.~\ref{fig3:jetmult}  is remarkable, particularly for the multiplicities larger than the maximum number of jets (three) for which the 
exact LO matrix-element calculation is performed. This underscores the potential benefit of the TMD evolution in better describing hard and non-collinear emissions, compared to the standard collinear evolution.

\begin{figure}[hbtp]
  \begin{center}
	\includegraphics[width=.23\textwidth]{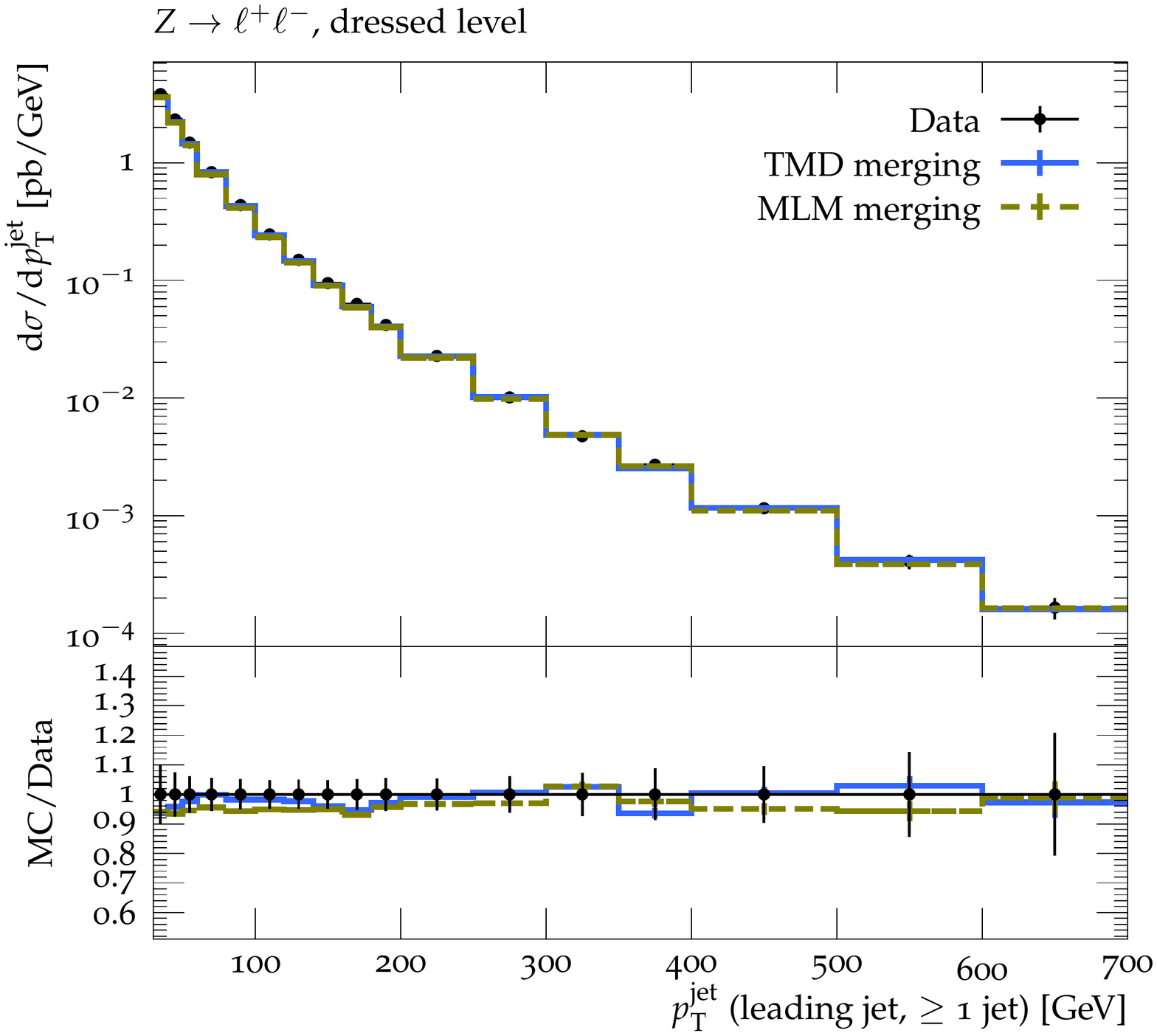}
	\includegraphics[width=.23\textwidth]{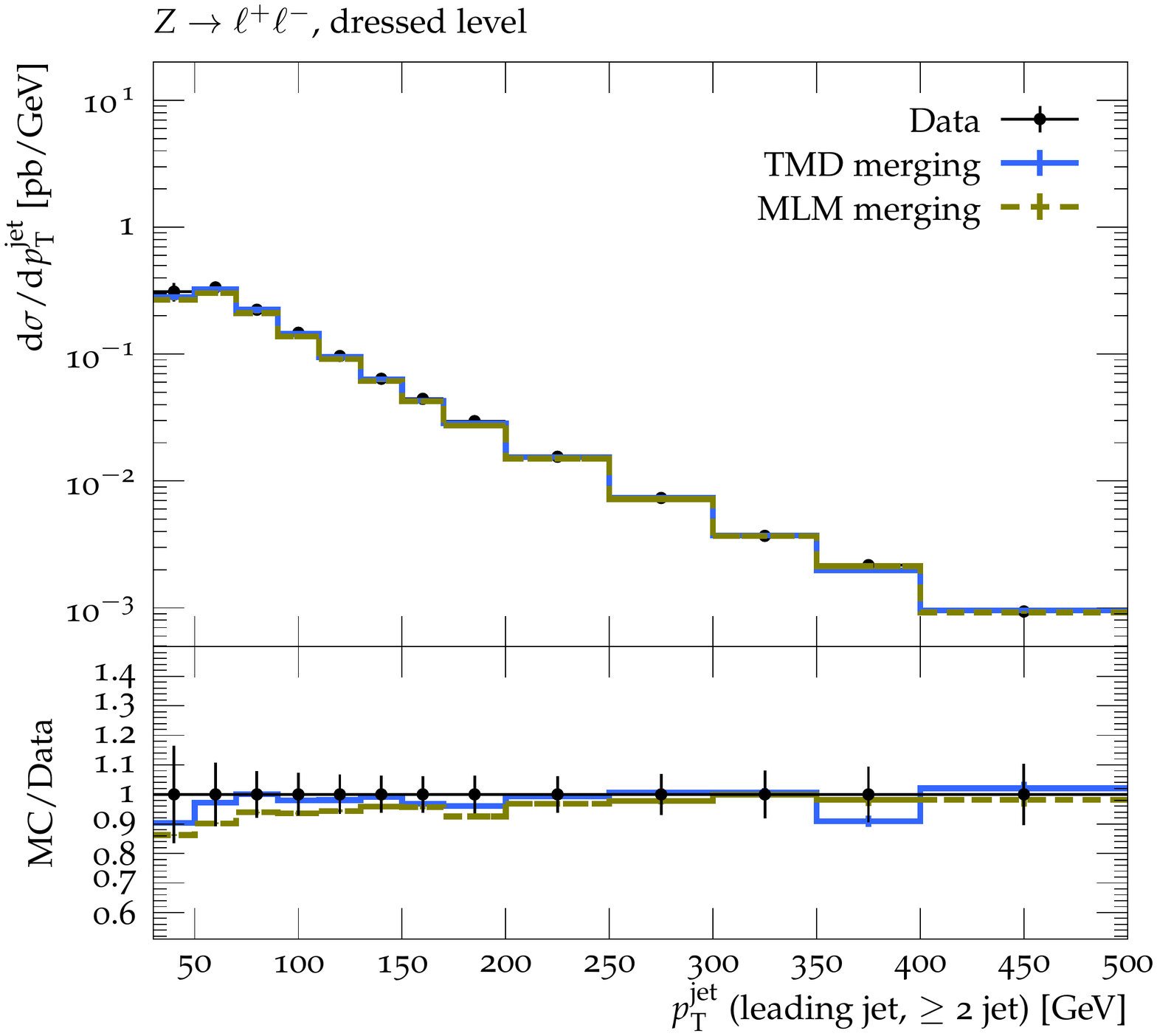}
	\includegraphics[width=.23\textwidth]{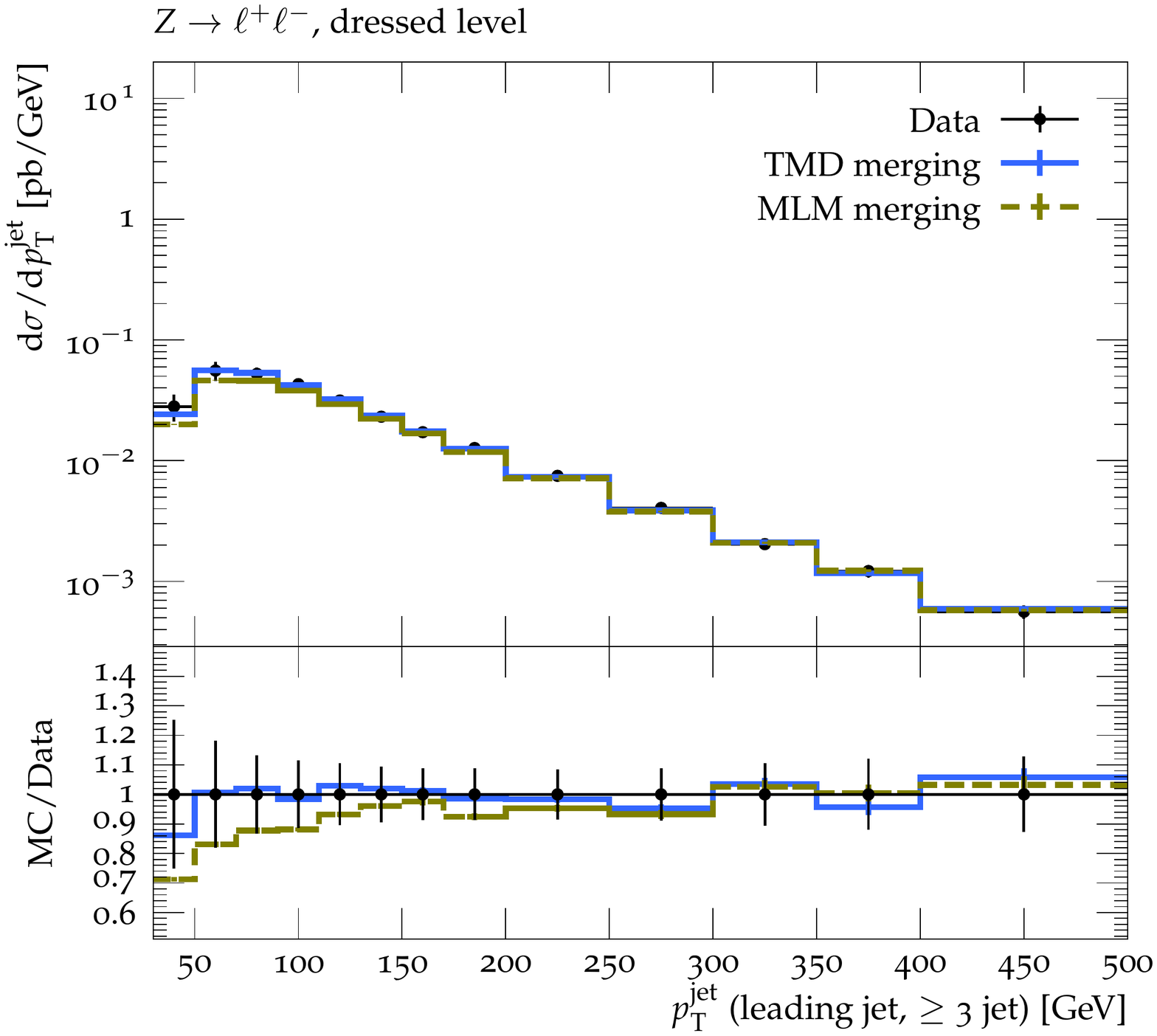}
        \includegraphics[width=.23\textwidth]{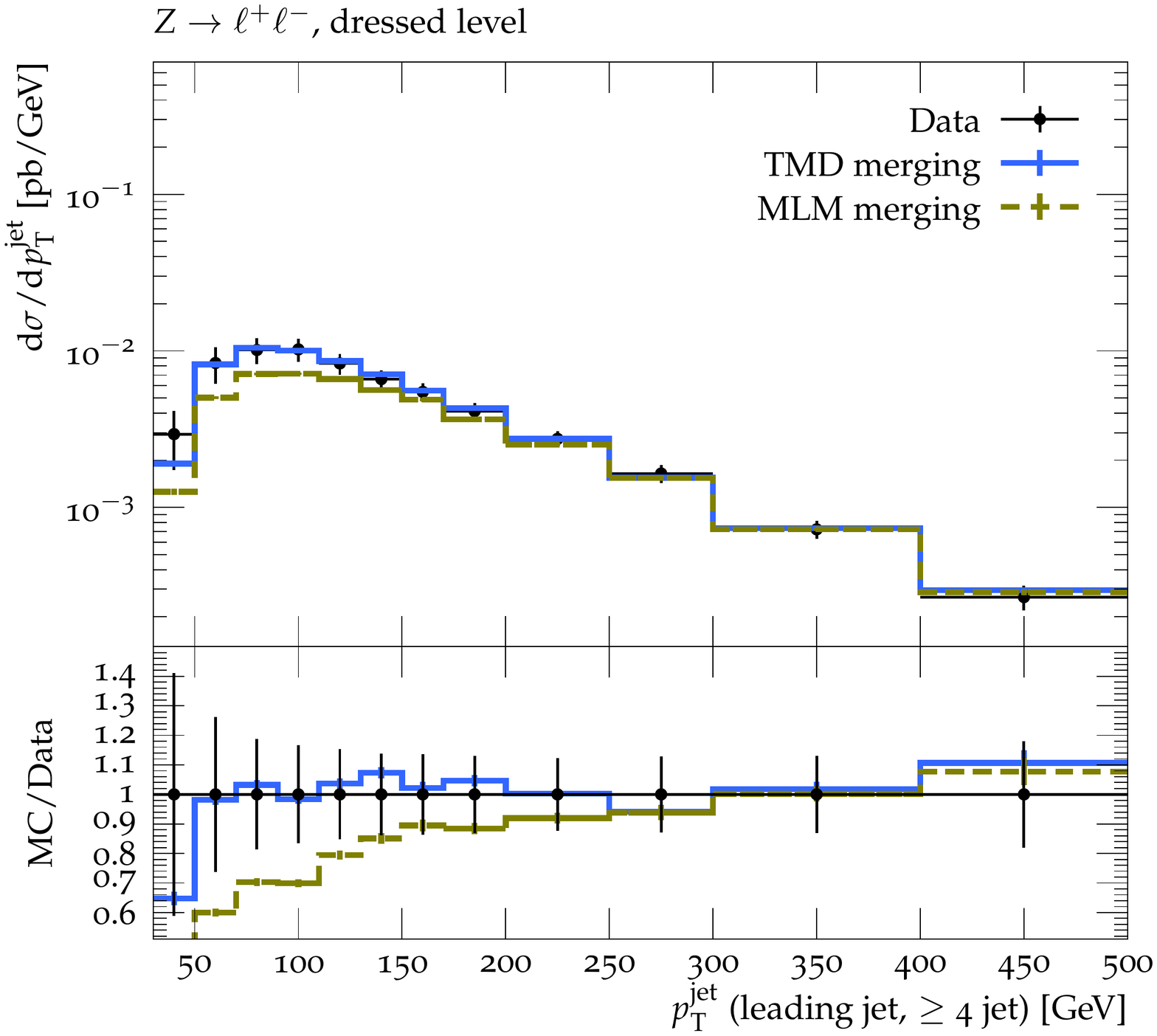}
  \caption{Leading jet $p_T$ spectrum in  inclusive production of $Z+$1 jet  (top left), $Z+$2 jets  (top right), 
  $Z+$3 jets  (bottom left), and    $Z+$4  jets (bottom right).   
  Experimental measurements by ATLAS~\protect\cite{Aaboud:2017hbk} at  $\sqrt{s} = 13$ TeV are shown. 
We plot the results of the two merging prescriptions, TMD (blue curves) and  MLM (green curves). The latter was obtained 
  using  {\sc Madgraph}+{\sc Pythia}6 with MLM merging.}	  
  \label{fig4:jetpT}
  \end{center}
\end{figure}

In Fig.~\ref{fig4:jetpT}   we investigate this further by examining the transverse momentum spectra of the associated jets.   
We  compare the results of the TMD calculation with the results from the collinear merging calculation which is obtained by 
 replacing the initial-state TMD shower evolution with collinear shower evolution implemented in {\sc Pythia}6,   
while keeping the same  matrix-element and final-state shower evolution parameters in the two calculations.  Clear differences emerge in 
the spectra that are most sensitive to higher-order shower emissions, such as the leading jet $p_T$ distribution in final states 
with at least 4 jets. The description of the jet $p_T$ improves thanks to TMD with respect to collinear merging at high multiplicities.

\begin{table}[htb!h]
  \centering
  \begin{tabular}{|c|c|c|c|c|c|}
    \hline\hline
       Merging  & $\sigma[\text{tot}]$ & $\sigma[\geq 1\text{ jet}]$ & $\sigma[\geq 2\text{ jet}]$ & $\sigma[\geq 3\text{ jet}]$ & $\sigma[\geq 4\text{ jet}]$ \\
       scale $[\text{GeV}]$ & $[\text{pb}]$ & $[\text{pb}]$ & $[\text{pb}]$ & $[\text{pb}]$ & $[\text{pb}]$ \\
    \hline \hline
    23.0 & 573 & 87.25 & 20.27 & 4.84 & 1.18 \\
    \hline
    33.0 & 563 & 86.15 & 20.48 & 4.86 & 1.19 \\
    \hline\hline
  \end{tabular}
  \caption{Multi-jet rates from the TMD merging algorithm as a function of the merging scale.}
  \label{tab1}
\end{table}

As mentioned earlier, the inclusion of transverse momentum recoils through TMD evolution can influence 
the theoretical systematics associated with the multi-jet merging algorithm when  matrix-element and parton-shower contributions are combined. Here we focus in 
particular on the systematic effects occurring through the dependence on the merging scale. 
In order to assess such systematics, we show  in Tab.~\ref{tab1} the multi-jet rates for different multiplicities  
for a 10 GeV variation of the merging scale. We observe that a 10 GeV variation of the merging scale results in less than 2\% variation for all the jet multiplicities 
considered. The results correspond to $pp$ collisions at 13~TeV, and jets are defined according to the selection and cuts used by ATLAS in Ref.~\cite{Aaboud:2017hbk}. Contrary to the results in previous figures, and to facilitate the comparison between different merging cuts, the rates shown here are absolute, i.e. they are not rescaled to the NNLO total cross section. The systematics is significantly smaller than what found with standard algorithms of collinear merging, as reported in Ref.~\cite{Alwall:2007fs}, where 
the variation of the jet multiplicity rates was found to be about 10\% when a 10 GeV change in the merging scale is considered.

In 
summary, we have presented a new multi-jet merging methodology  to analyze jet final states in high-energy hadronic collisions, 
by incorporating the physics of initial-state TMD evolution. With the high statistics available for precision measurements of fundamental interactions at forthcoming 
collider experiments,  this approach opens new avenues for   advanced studies of strong interaction dynamics in high multiplicity final states. In particular, 
while TMD effects have mostly been studied so far in the low-$p_T$ inclusive spectra,  our approach allows one to initiate investigations of possible TMD effects  
at the level of exclusive jet observables and in the domain of the highest $p_T$ processes, where the impact on  searches for signals of new physics may be largest. 

We have shown examples in $Z$+jets events in which contributions of 
higher jet multiplicities, undergoing  TMD evolution, are successfully merged via the new method. 
  The new method leads to lower variation with the merging scale of the individual multi-jet cross sections 
compared to non-TMD, collinear algorithms. 

We have presented predictions  for the $Z$-boson $p_T$ spectrum, for the associated 
jet multiplicity distributions and for the associated jet $p_T$ spectra, and compared them  
with experimental measurements at the LHC. We find very good agreement over a broad range in transverse momentum scales. This  points to the  
possibility of using the method uniformly from the lowest   to the highest $p_T$ accessible at current and future experiments, and underlines the 
relevance of further extending the merging approach with TMD evolution to NLO multi-jet samples.  

As the TMD broadening grows with 
increasing evolution scale $\mu$ and decreasing longitudinal momentum fraction $x$, we expect the 
effects studied in this paper to become even more relevant  in the case of the higher  scales probed at 
higher luminosity~\cite{Azzi:2019yne} and 
higher energy~\cite{Mangano:2016jyj} colliders.

{\bf Acknowledgments.} We are grateful to H.~Jung for useful discussions. FH thanks the Theory Department at CERN for hospitality and support 
while part of this work was being done.



\end{document}